\documentclass[12pt]{article}
\usepackage{epsfig,subfigure,}
\def\bkg{background}
\def\he4{ ^4\!{\rm He} }
\def\c14{ ^{14}\!{\rm C} }
\def\carb{ ^{12}\!{\rm C} }
\def\enu{E_{\nu}}
\def\Te{T_e}
\def\angsun{\theta_{\odot}}
\def\be7{^7\!{\rm Be}}

\def\WAYNE{$^a$}
\def\ARIZ{$^b$}
\def\CDF{$^c$}
\def\EFI{$^d$}
\def\PITT{$^e$}
\def\FSU{$^f$}

\begin{document}
\title{A Preliminary Look at the Physics Reach of a Solar Neutrino TPC: 
       Time-Independent Two Neutrino Oscillations}
\author{ G. Bonvicini\WAYNE, 
         E. Cheu\ARIZ, 
         J. Dolbeau\CDF, 
         P. Gorodetzki\CDF,  \\ 
         R. Kessler\EFI, 
         V. Paolone\PITT, 
         H. Prosper\FSU
             \vspace{0.5cm} \\
\WAYNE{\it Wayne State University, Detroit MI 48201}\\
\ARIZ{\it University of Arizona, Tucson AZ 85721}\\
\CDF{\it College de France, F-75231 Paris}\\
\EFI{\it Enrico Fermi Institute, University of Chicago, Chicago IL 60637}\\
\PITT{\it Univ. of Pittsburgh, Pittsburgh PA 15260}\\
\FSU{\it Florida State University, Tallahassee FL 32306}\\
}
\maketitle
\bigskip
\begin{center}{\it Contributed to Snowmass 2001, The Future of Particle Physics}\end{center}
\bigskip\par
\begin{center}
{\bf Abstract}
\end{center}
\vskip 0.1in

Solar neutrino physics below 2 MeV is a new frontier in particle
and astrophysics. More than 99\% of the solar neutrinos are produced in that
region, and the physics output of a successful experiment includes stringent
tests of stellar interior models. The original motivation to study solar
neutrinos 40 years ago was indeed the study of the Sun's interior.
Even more important, from a particle physics perspective, 
is the opportunity to study neutrino mixing using a well-calibrated
neutrino source that produces pure $\nu_e$ at $t=0$.
This high intensity solar neutrino source allows sensitivity
to neutrino masses down to $10^{-11}eV^2$, 
and it may produce enhanced sensitivity to neutrino mixing 
parameters via the strongly non-linear MSW effect.

This paper will discuss the physics reach of a solar neutrino TPC
containing many tons of $\he4$ under high pressure.
Particular attention is given to the LMA and SMA solutions,
which are allowed by current data\cite{bahcall}, 
and which are characterized by a lack of time-dependent phenomena (either
summer-winter or day-night asymmetries). 
In this case, the physics of
neutrino masses and mixing is all contained in the energy 
dependence of the electron neutrino survival probability, $P_{e}(E)$
(or in its reciprocal, the electron neutrino disappearance probability,
 $P_{x}(E)=1-P_e(E)$). 
While it is clear that the  $P_{e}(E)$ functions considered here cannot
be the correct ones, due to the presence of a third neutrino, the
physics reach of the TPC is analyzed within the context of the 
model to initiate a comparison between experiments.
In Section 1 the observables available to a TPC are discussed. 
Section 2 describes the simulation methods, analysis and 
input parameters, and illustrates some qualitative features
of the TPC capability using only the reconstructed neutrino energy.
Section 3 presents more advanced fits using both the recoil electron
energy and the reconstructed neutrino energy.

\newpage

\section{The information content of a TPC}

Experimentally one can in principle count four neutrino species below 2 MeV;
$pp$, $\be7$, $pep$ and $NO$, where the CNO spectra are counted as one.
For each species there is a $\nu_e$ flux $\Phi_e$,
a $\nu_x$ flux $\Phi_x$ ($x=\mu,\tau$), 
and at least one energy dependent parameter
describing the change of $P_{e}(E)$ across the spectrum of the species.
\begin{figure}[htb]
 \begin{center}
    \mbox{\epsfysize8cm\epsffile{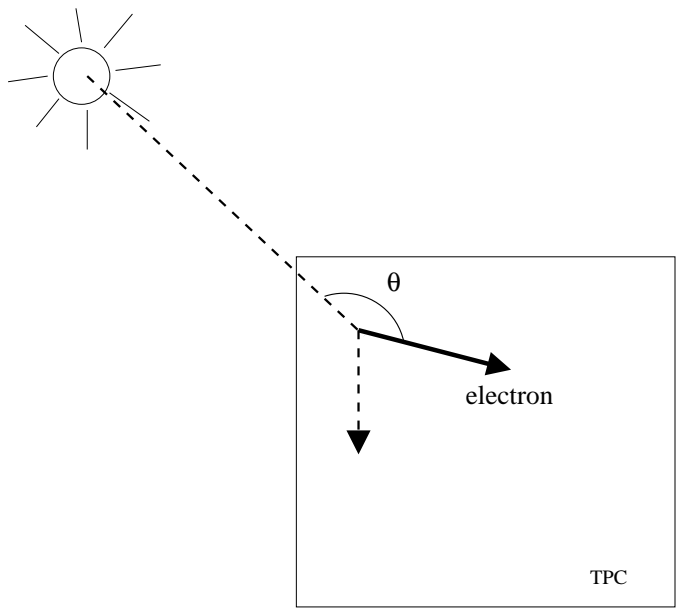}}
    \caption{ The solar neutrino detection scheme. The recoiling
electron energy and direction are measured, and the neutrino energy
is reconstructed according to Eq. 1. Only a small fraction of the solid angle
around the solar axis is considered at any given time.
\label{fn:cartoon} }     
 \end{center}
\end{figure}

In our approach, the scattering of a solar neutrino off of a target electron
($\nu e\to \nu e$)
results in a completely reconstructed electron track
(Fig.\ref{fn:cartoon}). From the simultaneous reconstruction
of the electron recoil energy ($\Te$) and its angle with respect 
to the solar axis ($\angsun$),
the neutrino energy for each event is given by
\begin{equation} 
   \enu = {m\Te\over p\cos{\angsun} - \Te} ~ .
\end{equation}
This results in a direct measurement of the neutrino energy spectrum,
and can be compared with solar models combined with neutrino mixing models.
The correlations between $\Te$ and $\enu$ can also be used to 
measure the neutrino flavor content (see Sec.~3).
This method works best when $\enu \sim m$. At energies far higher
than the electron mass the error propagation becomes unfavorable.

A TPC is currently being designed to measure electron tracks for
$\Te > 100$~keV. The lowest energy $100$~keV tracks travel several cm 
before stopping, providing adequate track lever arm. 
The $\nu e$ elastic scattering cross-section is known
to better than 1\%, which effectively eliminates any theoretical error 
on the solar flux measurement.
Further, the effective imaging of the event
provides many opportunities for precision calibration of the device.
It is very important to know the detector resolution
function precisely. 

The two main kinematically constrained types of events that one can use 
to study the detector (there are several types one can use) are $\delta-$rays
from cosmic rays and double events. Every ten cosmic rays will produce
a $\delta-$ray with a kinetic energy $\Te$ in excess of 100 keV, 
and distributed like $1/\Te^2$. 
If the cosmic is extremely relativistic, and typical energies
underground average 300 GeV, then the following relation between the 
electron kinetic energy, electron mass, and its angle with respect 
to the cosmic direction is
\begin{equation}
  \Te = 2m/\tan^2\theta_e ~.
\end{equation}
Because generally $\Te < m$, these electron tracks are at large angles
and can be easily separated from the column of ionization from the 
cosmic ray.
At 2500 mwe\footnote{mwe = meter-water-equivalent}
we expect $\sim 10^6$ such events per year, providing
virtually unlimited tagged, quality calibration events.

\begin{figure}[htb]
 \begin{center}
    \mbox{\epsfysize8cm\epsffile{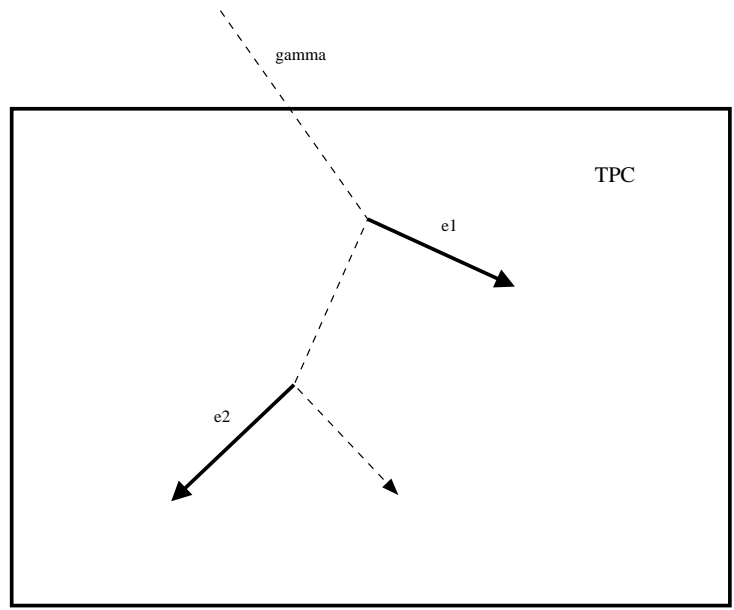}}
    \caption{ Double-Compton background events in the TPC. 
              A gamma ray enters the TPC and scatters twice, 
              producing two usable electron tracks. 
\label{fn:cartoo2} }     
 \end{center}
\end{figure}

Another type of calibration event, ``double-Compton'', 
consists of a photon that 
Compton-scatters twice in the TPC and produces two usable tracks 
which are correlated in time and angle (see Fig.\ref{fn:cartoo2}).
Eight quantities are recorded by the TPC (3-momentum of each electron
and direction between two vertices)
against five possible degrees of freedom. In the simplest possible 
application the intermediate photon (the one connecting the two
electron tracks)
is reconstructed at both vertices using Compton kinematics. 
The exact kinematic relation
\begin{equation}
   {p_1\cos\theta_1\over T_1} +{p_2\cos\theta_2\over T_2} = 2 ~, 
   \label{eq:dcompton}
\end{equation}
can be compared with the measured quantities to provide
an accurate energy calibration.
The angles $\theta_{1,2}$ are with respect to the direction of the
intermediate photon. With the background conditions described below, 
$\sim 10^4$ double-Compton events per year are expected

The calibration is expected to be of such quality that the energy scale 
and resolution will be known to $\sim 1$\%\cite{arzarello}. 
These techniques can also map out position and time-dependent 
non-uniformities in the detector response.
Many simulations have demonstrated that the reconstructed neutrino energy 
resolution is dominated by the angular resolution of the electron track.
A gas mixture which is Helium dominated provides very low multiple
scattering, as well as the opportunity to cold-trap impurities.

The other fundamental reason to have directionality is, of course, the
direct measurement of the backgrounds. Except for the 6.8\% variation in
the flux due to the Earth eccentricity, our neutrino source varies only
in direction. As abundantly proven by Kamiokande, directionality allows
precise measurements in the presence of relatively high backgrounds,
and will be demonstrated in greater detail later in this study.
In practice the detector is sensitive
only to the statistical fluctuations in the background, and the statistical
significance ultimately will depend on $S/\sqrt{B}$.

\clearpage
\section{Simulated Energy Spectra and Backgrounds}

\begin{figure}[htb]
 \begin{center}
    \mbox{\epsfysize8cm\epsffile{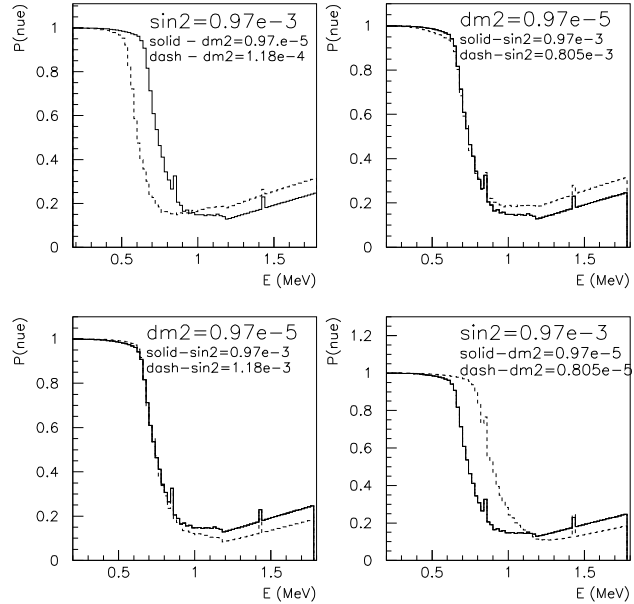}}
    \caption{ The neutrino survival probability in the SMA region, as a 
function of neutrino energy. The solid line is always the same, and represents
the calculation for the nominal point at $(\Delta m^2=0.97\times 10^{-5},
\sin^2{2\theta}=0.97\times 10^{-3})$. The dashed lines represent points
displaced from the nominal point by 21\%, respectively above, to the left,
to the right, and below the nominal point.
\label{fn:survpr0} }     
 \end{center}
\end{figure}

\begin{figure}[htb]
 \begin{center}
    \mbox{\epsfysize8cm\epsffile{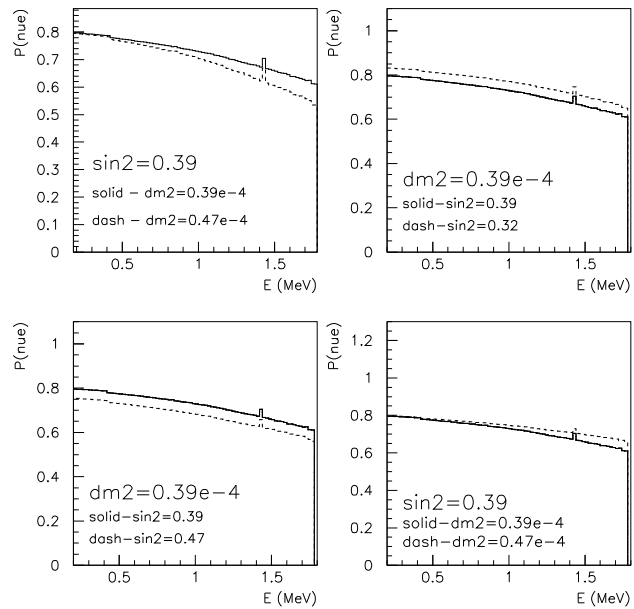}}
    \caption{ The neutrino survival probability in the LMA region, as a 
function of neutrino energy. The solid line is always the same, and represents
the calculation for the nominal point at $(\Delta m^2=3.9\times 10^{-5},
\sin^2{2\theta}=0.39)$. The dashed lines represent points
displaced from the nominal point by 21\%, respectively above, to the left,
to the right, and below the nominal point.
\label{fn:survpr1} }     
 \end{center}
\end{figure}

The neutrino survival probabilities were computed by using the analytic
formulae of Ref.~\cite{haxton}, convoluting them over the neutrino
production region and propagating them through solar matter with the
density profile described by the SSM.
The MSW-SMA region is, generally speaking, characterized by a strongly varying
$P_{e}(E)$. The MSW-LMA region has a slow, 
continuous spectral distortion across the 0-2 MeV region. 
Fig.~\ref{fn:survpr0} shows $P_{e}(E)$ for a conveniently chosen
central point in the SMA region, and for four points nearby corresponding
to a 21\% variation up and down in both mass and 
$\sin^2{2\theta}$. 
Fig.~\ref{fn:survpr1} does the same for the LMA region. 
Because no time-dependent effects are
expected in this region, the most general analysis is done by studying
the $(\enu,x)$ scatter plot, where $x=\Te/\Te^{max}$.

Monte Carlo events were generated using the parameters listed in 
Table \ref{tn:Machines}. 
Results are presented for exposures of 7 and 70 Ton-years, roughly 
corresponding to one and ten years of data taking. 

\begin{table}
\caption{Parameters used in the simulations described in the text.
\label{tn:Machines}}
\begin{center}
\begin{tabular}[t]{|c|c|}
\hline
  Parameter   & Value \\
 \hline
Exposure             & 7 \& 70 ton-years        \\
$\sigma_T/T$         & 0.05 at 100 keV          \\
$\sigma_\theta$      & $15^{\circ}$ at 100 keV  \\
$^{238}U$            & $0.5~\mu$g               \\
$\c14/\carb$         & $5\times 10^{-20}$g/g    \\
\hline
\end{tabular}
\end{center}
\end{table}

Substantial electron-track simulations have been undertaken 
over the years\cite{cdf}, which include the effects of 
multiple-scattering and diffusion in a large static electric field.
Amongst the recent results are the good agreement
of various simulation packages (GEANT and FLUKA), and an estimate
of the angular resolution which scales with pressure as $\sim 1/P^2$.
A complete optimization of the tracking algorithm represents a genuinely
new problem and will be part of our future R\&D.
Based on the results from these simulations,
we use the following detector resolutions:
\begin{equation}
   { {\sigma_T}\over T} (T)= { 0.016\over \sqrt{T} } ~~~~~~~~~~~~~~
   \sigma_\theta(T)={4.7^{\circ}\over T^{0.6}}
\end{equation}
where $T$ is the electron kinetic energy in MeV.
At the lowest track energy of 100~keV, this corresponds 
to resolutions of 5\% for energy and $15^{\circ}$ for angle.

The backgrounds were simulated as follows:
\begin{itemize} 
  \item $0.5~\mu$g of $^{238}U$ in secular equilibrium with its decay 
        products, and uniformly distributed on the inner TPC 
        surface. For each $^{238}U$ decay, this results in 
        0.68 $\gamma$'s  from $^{214}Pb$ and 
        1.31 $\gamma$'s  from $^{214}Bi$~; 
        the total number of emitted $\gamma$'s is $4.3\times 10^5$ 
        for 7 ton-years.
 \item  5$\times 10^{-20}$ $^{14}C/^{12}C$ in the methane quencher,
        which corresponds to a $\c14/\he4$ ratio of $5\times 10^{-22}$.
\end{itemize}
Further sources of background, such as 
\begin{itemize} 
\item $^{14}C$ in the TPC body (contributing
both direct tracks and bremstrahlung photons),
\item cosmic-generated backgrounds
\end{itemize}
have been discussed elsewhere\cite{arzarello}
and found to be negligible. Radon contamination is assumed to be avoidable,
because the helium methane mixture  
can be processed through a cold trap. This detector needs less than
a few $\mu Bq/m^3$, whereas purities an order of magnitude lower have
been achieved, for example by GNO\cite{heusser}. The radon-trapping efficiency
has been measured to scale approximately like $1/T^2$\cite{heusser}, 
implying a factor
of about two loss in efficiency when operating the trap at the boiling point
of methane, which is 112K.

\clearpage

\begin{figure}
  \centering \epsfig{file=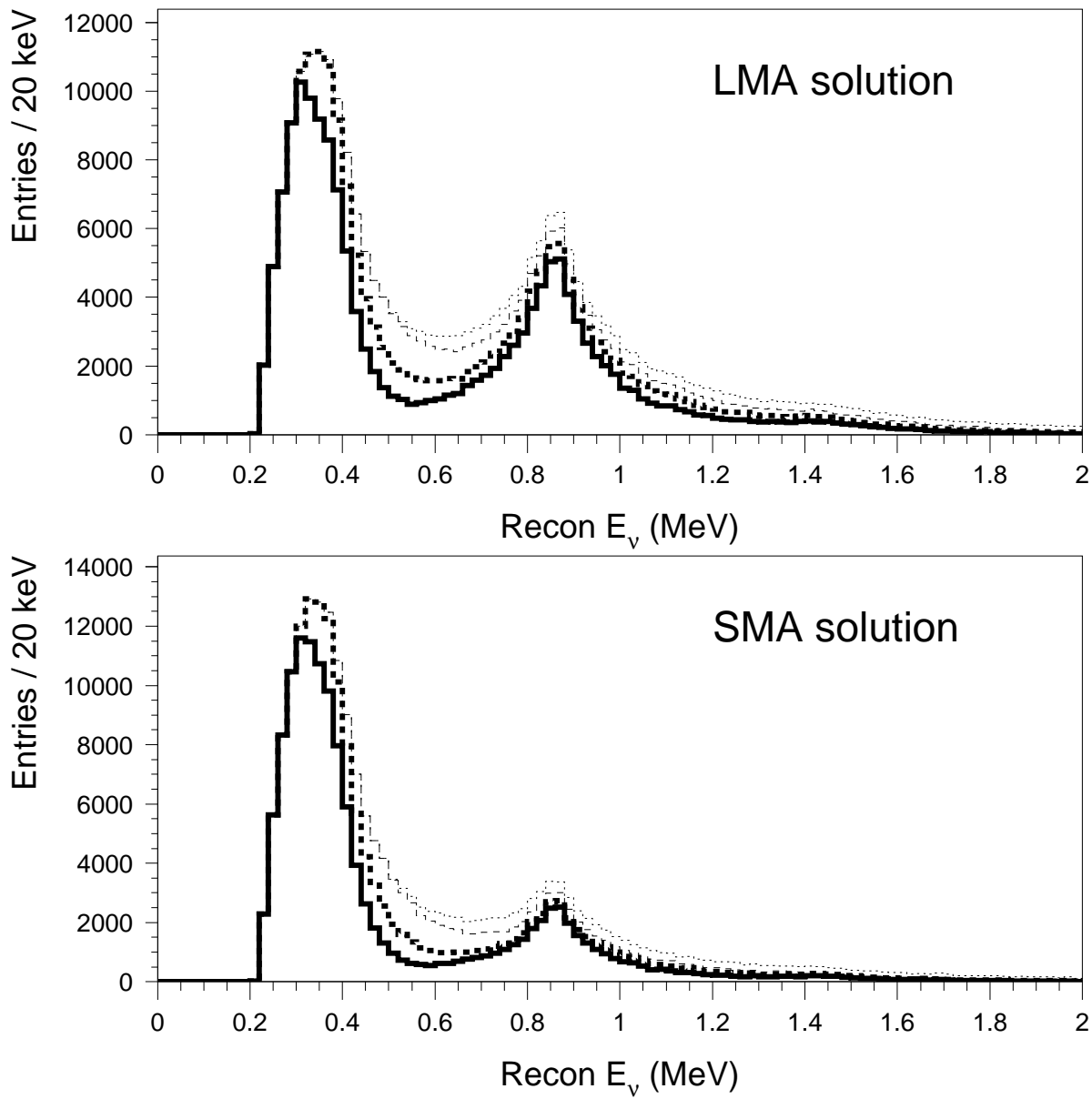,height=7.0in}
  \caption
  { \footnotesize  
       Signal neutrino energy spectra (no {\bkg}-subtraction)
       with different cuts on the cosine of the angle relative to the sun;
       no cut (dot-dash), 
       $\cos\angsun > 0.6$ (dash), 
       $\cos\angsun > 0.7$ (thick-dash), 
       $\cos\angsun > 0.8$ (thick-line). 
  }
  \label{fig:nuspect thetacut} \end{figure}

The expected rates in the TPC detector are shown in 
Table~\ref{tb:nurates}.  The kinematic cut $\cos\angsun > 0.8$
retains 2/3 of the signal and improves the separation between
the $pp$ and $\be7$ peaks (see  Figure~\ref{fig:nuspect thetacut}),
and it also rejects 1/2 of the $U$-decay
and 2/3 of the $\c14$ background. The remaining {\bkg}-rejection
comes from requiring $T_e > 0.1$~MeV and that the reconstructed
neutrino energy is above 0.2~MeV. 
The typical signal rates are 1-2$\times 10^3$ per 7 ton-years,
depending on the neutrino mixing solution, 
and the signal/{\bkg} ratio is of order unity.

\begin{figure}
  \centering \epsfig{file=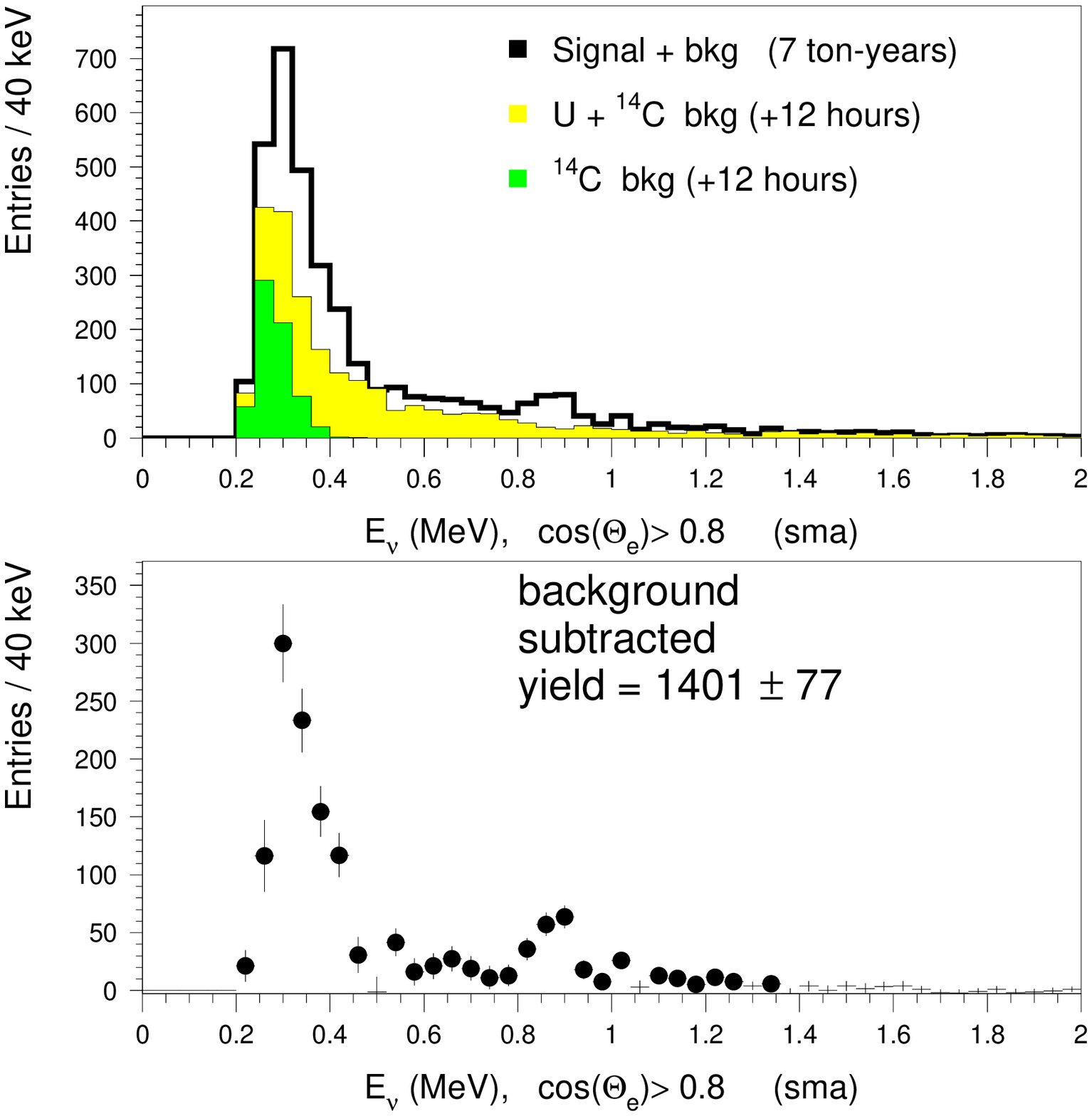,height=7.0in}
  \caption
  { \footnotesize  
       MC prediction for experimental neutrino energy spectrum based on
       {\bf 7} ton-years and with SMA solution 
       ($\Delta m^2 = 0.97\times 10^{-5}~{\rm eV}^2,~
         {\sin}^2 2\theta=0.00097$).
       Top plot shows total spectrum with {\bkg s}, and bottom
       plot shows spectrum obtained after subtracting {\bkg} based
       on events collected 12 hours out of phase.
  }
  \label{fig:nuspect sma07} \end{figure}

\begin{figure}
  \centering \epsfig{file=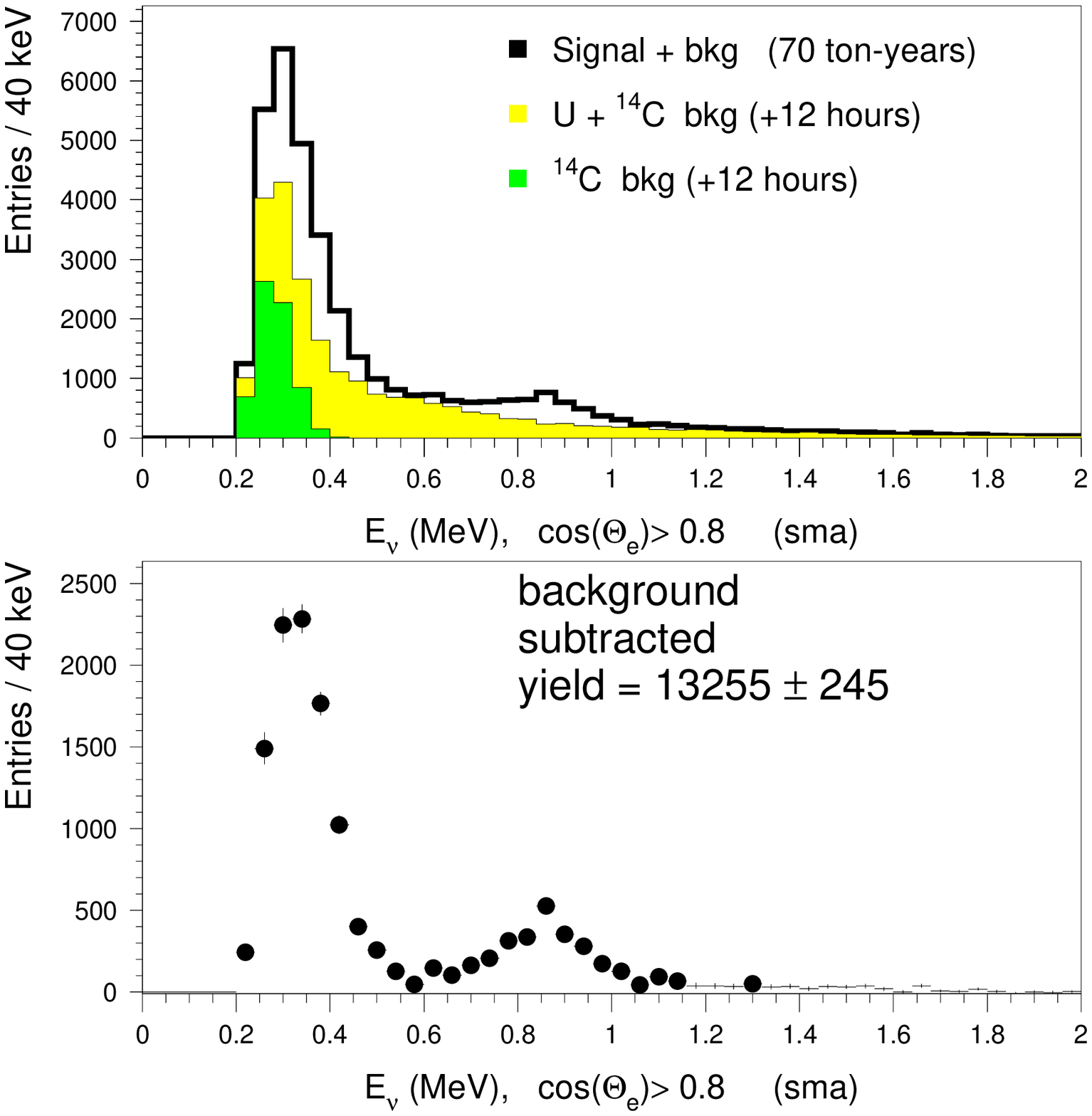,height=7.0in}
  \caption
  { \footnotesize  
       MC prediction for experimental neutrino energy spectrum based on
       {\bf 70} ton-years and with SMA solution 
       ($\Delta m^2 = 0.97\times 10^{-5}~{\rm eV}^2,~
         {\sin}^2 2\theta=0.00097$).
       Top plot shows total spectrum with {\bkg s}, and bottom
       plot shows spectrum obtained after subtracting {\bkg} based
       on events collected 12 hours out of phase.
  }
  \label{fig:nuspect sma70} \end{figure}

Using the SMA solution, an MC prediction of the ``experimental'' 
neutrino energy spectrum is shown in 
Figures~\ref{fig:nuspect sma07}-\ref{fig:nuspect sma70}
for 7 and 70 ton-years.  The top plots show signal+{\bkg},
and the shaded overlays show the {\bkg} predictions based on
events that point $180^{\circ}$ away from the sun; 
i.e., 12 hours out of phase. The bottom plots show the
resulting $\enu$ ``physics'' spectrum after {\bkg}-subtraction.

\begin{figure}
  \centering \epsfig{file=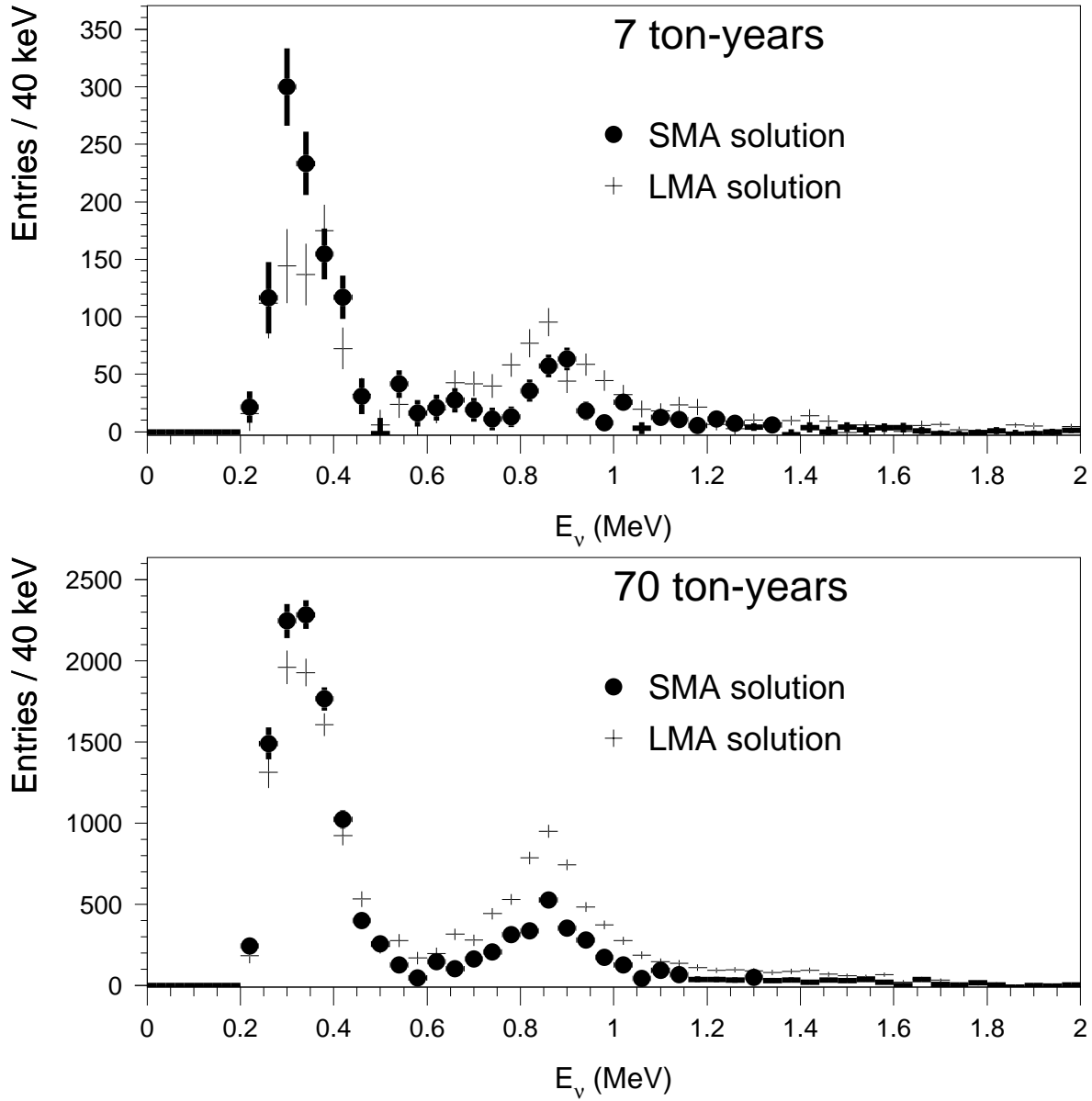,height=7.0in}
  \caption
  { \footnotesize  
       MC prediction of experimental neutrino energy spectrum for 
       LMA (dots) and SMA (crosses) solutions. 
       The top (bottom) plot corresponds to 7 (70) ton-years.
  }
  \label{fig:nuspect lmasma} \end{figure}

\begin{figure}
  \centering \epsfig{file=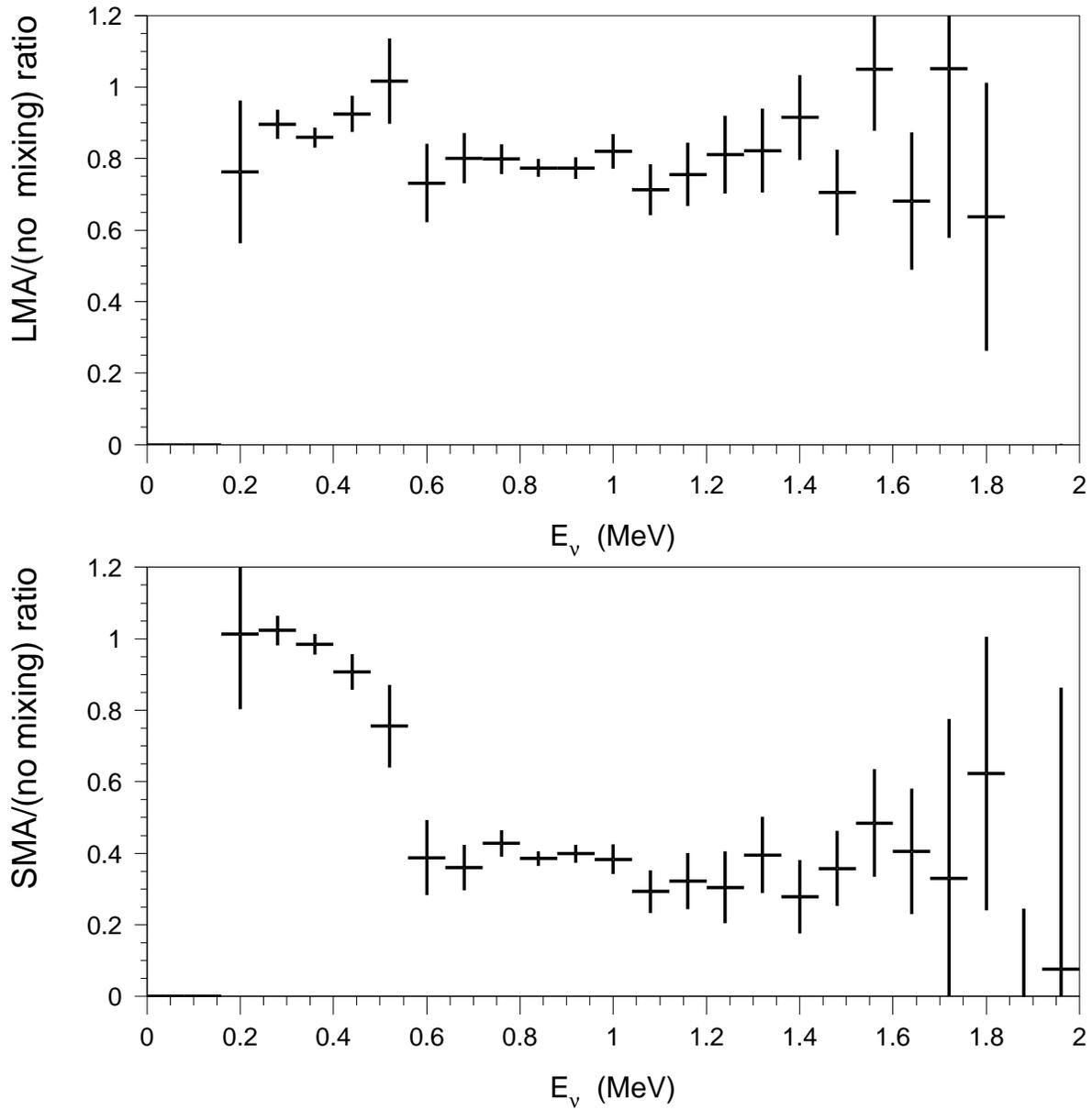,height=7.0in}
  \caption
  { \footnotesize  
       Ratio of predicted LMA $\enu$ spectrum to spectrum with 
       no-mixing (top) and same for SMA solution (bottom).
       The statistical uncertainties are for 70 ton-years.
  }
  \label{fig:nuspect ratio} \end{figure}

\begin{figure}
  \centering \epsfig{file=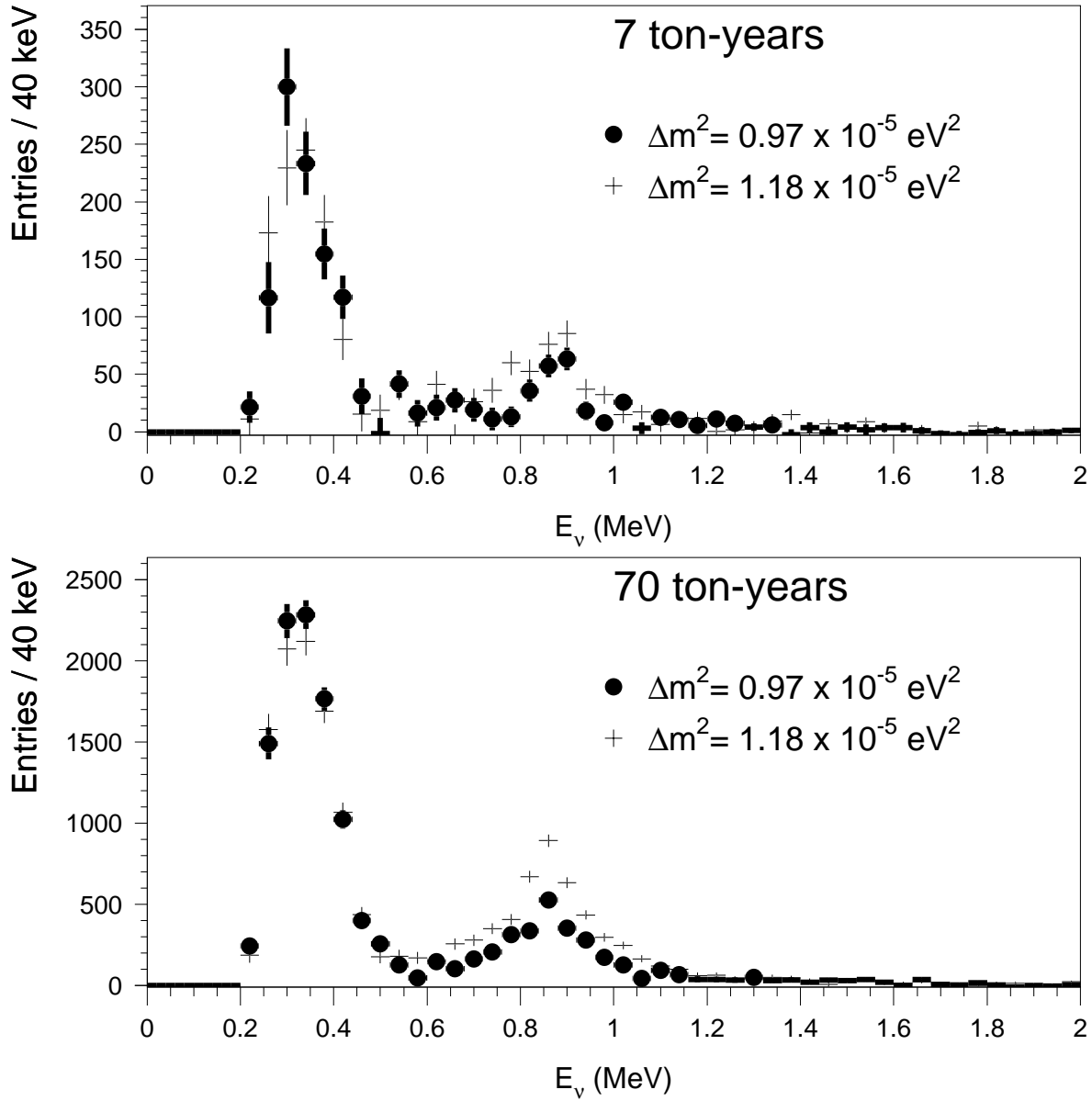,height=7.0in}
  \caption
  { \footnotesize  
       MC prediction of experimental neutrino energy spectrum for 
       two SMA solutions with similar $\Delta m^2$ values as
       indicated on the plots
       (${\sin}^2 2\theta = 0.00097$ for both).
       The top (bottom) plot corresponds to 7 (70) ton-years.
  }
  \label{fig:nuspect smasma} \end{figure}

The TPC measurement capability is illustrated in
Figures~\ref{fig:nuspect lmasma}-\ref{fig:nuspect smasma}.
Figure~\ref{fig:nuspect lmasma} compares the SMA and LMA
solutions; there is some distinction with 7 ton-years
and a very clear distinction with 70 ton-years.
Additional information in $\enu$ vs. $T_e$ correlations
give additional separation (see Sec.~3).
Figure~\ref{fig:nuspect ratio} shows the ``experimental''/no-mixing 
ratio vs. $\enu$ for the LMA and SMA solutions. Note that
$\nu_{\mu}e\to \nu_{\mu}e$ scattering in the TPC results in ratios
that are slightly different than the $\nu_e$ survival
probabilities in Figures~\ref{fn:survpr0}-\ref{fn:survpr1}.

Figure~\ref{fig:nuspect smasma} compares two SMA solutions
with slightly different values of $\Delta m^2$.  The sensitivity
to $\Delta m^2$ depends on the particular values
of the neutrino mixing parameters, so the resulting uncertainty
could be better or worse than the 20\% $\Delta m^2$-difference shown in
Figure~\ref{fig:nuspect smasma}.

\begin{table}[hbt]
\centerline{ \parbox{6in}{
\caption[ ]
    { 
       Rates per 7 ton-years for different neutrino mixing models
       and for {\bkg s}.  ``RAW'' corresponds to electron events in the 
       detector with no kinematic cuts. The ``RAW U-decay+Compton'' entry
       reflects $\gamma$'s that have Compton-scattered in the TPC,
       which is 7\% of all $\gamma$'s from the U-decay chain.
       The rates are for  $0.2 < \enu < 2.0$~MeV, 
       where $\enu$ is the reconstructed neutrino energy.
       The LMA solution corresponds to
           $\Delta m^2 = 3.9\times 10^{-5}~{\rm eV}^2,~
             {\rm sin}^2 2\theta=0.39 $;
       the SMA solution corresponds to
           $\Delta m^2 = 0.97\times 10^{-5}~{\rm eV}^2,~
             {\rm sin}^2 2\theta=0.00097$.
    }
\begin{center}
\begin{tabular}{ | l | c c | } \hline
 Neutrino model or      &  \multicolumn{2}{c|}{Rate per 7 ton-years:}   \\
 {\bkg} source          &  RAW    &   after cuts   \\ \hline \hline
  SSM, no $\nu$ mixing  &  $2.98\times 10^3$   & $1.91\times 10^3$   \\
  SSM + MSW LMA         &  $2.45\times 10^3$   & $1.57\times 10^3$   \\
  SSM + MSW sma         &  $2.04\times 10^3$   & $1.31\times 10^3$   \\ 
 \hline
  $U$-decay + Compton   &  $3.27\times 10^4$  & $1.67\times 10^3$   \\
  $\c14~~~\beta$-decay  &  $6.40\times 10^4$  & $0.66\times 10^3$   \\ 
 \hline
\end{tabular}
\label{tb:nurates}\end{center}   }} \end{table}


\clearpage

%

\clearpage
\section{Preliminary $(E,T)$ analysis}

\def\enu{E_{\nu}}
\def\Te{T_e}


\def\mTEnu{ {m\Te} \over {\enu^2} }

The number of measured events in \{$\enu,~\Te$\} bins is related 
to the $\nu e$ scatter cross-sections by
\begin{eqnarray}
  { {d^2 N} \over {d\enu d\Te} } & = &
       P_e {d\sigma_{\nu_e e} \over d\Te} +
       (1-P_e) {d\sigma_{\nu_{\mu}} \over d\Te}    \\
      &   &  \\
      & = & F_1 - s(1-P_e)F_2
\end{eqnarray}
where
\begin{eqnarray}
   F_1(\enu,\Te) & = & (1/2+s)^2+s^2(1-\Te/\enu)^2 - 
                        s(1/2+s)\left( 1-{\mTEnu} \right) \\
   F_2(\enu,\Te) & = & 2 - {\mTEnu}
\end{eqnarray}
and $s=\sin^2\theta_W$.

%
%
%
%
%

Flavor separation by this method works best for the low energy $pp$
neutrinos. It works worst for $\be7$ neutrinos, because at those energies
the two recoil spectra are very similar in shape. 
The strength of this method of analysis is in the model independent
measurements that can be obtained by observing, on a statistical
basis, both electron and non-electron neutrinos.

The elastic scatter cross section $\sigma(\nu_x e\to \nu_x e)$
is roughly 1/4 of $\sigma(\nu_e e\to \nu_e e)$,
and the ratio varies slowly with energy. A change in $P_{e}$ with the
energy $E$ is always accompanied by a change in the slope along $T$.

As mentioned above, if there is no observable time dependence in the solar
neutrino flux, then all the information is obtained from the study of
the scatter plot of any two quantities $\angsun$, $T$ and $E$.
Monte Carlo events were generated with large statistics (7000 Ton Years exposure,
which is assumed to generate a negligible statistical error compared to 
the exposure of a real life experiment)
on a fine grid (21$\times$21 points, with step size between 5\% and 15\%) on the 
$(\Delta M^2,\sin^2{2\theta})$ plane. For each point on the grid, the $(E,T)$
distribution was recorded.

A further set of Monte Carlo events was generated, to be divided in simulated
experiments of the appropriate exposure (7 or 70 ton Years). A binned
likelihood function in $(E,T)$ 
was then used to compare the experiment under consideration
with each point in the lattice, and the one with the best $\chi^2$ was found.

This method is rather simple, but it conveys an estimate of the physics reach of 
the detector when the recoil information is included. At the same time, it avoids
the need to find an analytic fitting function, and does not need a parabolic
$\chi^2$ (nearly impossible, given the strong non-linearity of the MSW effect)
to converge.

\begin{figure}[htb]
 \begin{center}
    \mbox{\epsfysize7cm\epsffile{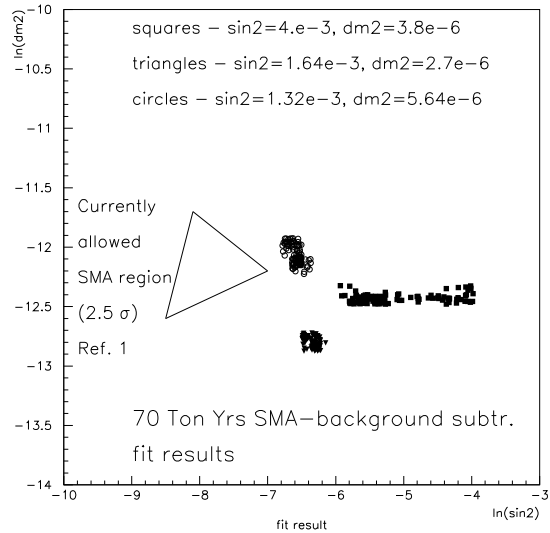}}
    \caption{ SMA region: The result of 100 fits at three different
values  in ($\Delta M^2,\sin^2 2\theta$) using a statistics equivalent to
70 Ton-years of exposure. Backround subtraction as discussed in the
text. The axes
represent the logarithm of the usual $(\sin^2{2\theta},\Delta M^2)$ values.
The currently allowed region is drawn for comparison.
\label{fn:fitssma} }     
 \end{center}
\end{figure}

\begin{figure}[htb]
 \begin{center}
    \mbox{\epsfysize7cm\epsffile{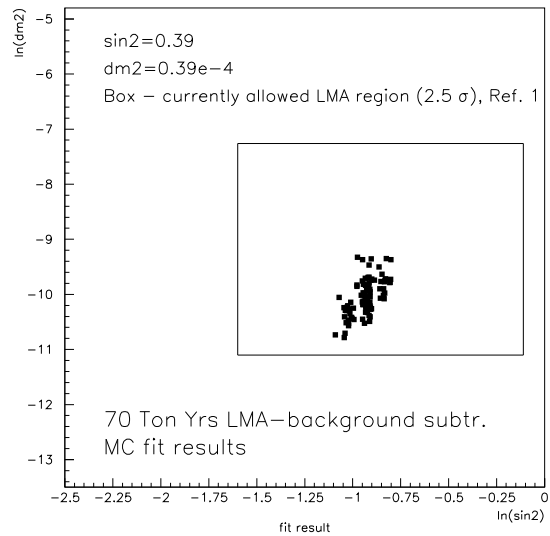}}
    \caption{ LMA region: 
The result of 100 fits using a statistics equivalent to
70 Ton-years of exposure. Background subtraction as discussed in the
text have been subtracted. The axes
represent the logarithm of the usual $(\sin^2{2\theta},\Delta M^2)$ values.
The currently allowed region is drawn for comparison.
\label{fn:fitslma} }     
 \end{center}
\end{figure}

Figs. \ref{fn:fitssma} and \ref{fn:fitslma} 
show the scatter of the best-fit points (100 simulated fits each),
for a simulated 
exposure of 70 Ton Years, and background subtraction as described above.
Drawn on the figure is the rough size of the currently 99\% allowed LMA and SMA 
regions\cite{bahcall}.

From the figures, one can infer the expected final error for this experiment,  
listed in Table\ref{tn:reduct}. Also listed is the approximate reduction factor in the
$(\Delta M^2,\sin^2{2\theta})$ log-log allowed region, compared to the
current allowed regions.

\begin{table}
\caption{Physics reach of the TPC for an exposure of 70 Ton Years. All errors in
percent of the central value.
\label{tn:reduct}}
\begin{center}
\begin{tabular}[t]{|c|c|}
\hline
  Parameter   & Value(\%) \\
 \hline
$\delta(\Delta M^2)$, LMA & $\sim$ 30 \\
$\delta(\Delta M^2)$, SMA & 3 to 10 \\
$\delta(\sin^2 2\theta)$, LMA & $\sim$ 7 \\
$\delta(\sin^2 2\theta)$, SMA & 5 to 100 \\
Reduction factor, LMA & $\sim 20$ \\
Reduction factor, SMA & NA \\

\hline
\end{tabular}
\end{center}
\end{table}


Several comments are in order:
\begin{itemize}
\item according to Figs. 11 and 12, this detector would identify the correct
region. In the LMA case, it would reduce the allowed region by
about a factor of twenty. In some cases, some of 
the mixing parameters can be determined at 
the several percent level. 
\item this detector would allow extremely stringent checks of the sterile
neutrino hypothesis. By measuring all types of neutrinos, in all flavors,
it can compare directly against the solar luminosity to check that
all solar energy is produced in association with neutrinos.
\end{itemize}

\clearpage
\section{Conclusions}

With strong evidence for long distance neutrino mixing\cite{sno}, 
the next generation of experiments should aim for
precision measurements of neutrino mixing parameters.
The TPC detector described in this paper has the potential to
dramatically reduce the  allowed parameter space and to
provide stringent test of the Standard Solar Model.

\end{document}